# A New Hybrid Precoding Approach for Multi-user Massive MIMO over Fading Channels

Azadeh Pourkabirian, *Member, IEEE,* Kai Li, *Senior Member, IEEE,* Photios A. Stavrou, *Senior Member, IEEE,* Wei Ni, *Fellow Member, IEEE.*

*Abstract—* Hybrid precoding is an indispensable technique to harness the full potential of a multi-user massive multiple-input, multiple-output (MU-MMIMO) system. In this paper, we propose a new hybrid precoding approach that combines digital and analog precoding to optimize data transmission over multiple antennas. This approach steers signals in specific directions, leading to maximizing sum-rate and suppressing side-lobe interference. When dealing with complex signals, changes in phase are naturally associated with changes in angle, and these variations are inherently correlated. The correlation between the angle and phase is essential for accurately determining the channel characteristics. An important aspect of this approach is that we model the angle and phase as correlated variables following a bivariate Gaussian distribution, and for the first time, we define a joint angle and phase entropy to measure the uncertainty of angle and phase variations in wireless channels. This entropy is crucial to adapt the proposed precoding method with variations. Simulation result validate the accuracy of our analytical findings, demonstrating 18.31% increase in sum-rate and an 11.47% improvement in robustness compared to other state-of-the-art methods.

## I. INTRODUCTION

Multiuser multiple-input multiple-output (MU-MMIMO) systems [1] with an extremely large number of users/antennas are one of the potential communication paradigms to enable massive connections for a wide range of smart applications such as holographic projection, virtual reality, and robotic surgery [2],[3]. As one of the potential technologies, antenna arrays [4],[5] consisting of many radiators and elements with diversity gain are considered to be a promising solution to increasing capacity and data rate and improving the performance of MU-MIMO systems. However, hardware requirements and computational complexity (e.g., the precoding matrix and the channel estimation calculation) rise with antennas [6]. On the other hand, activating hundreds of antennas at the same time may not be energy efficient, since selecting more antennas demands more power for antenna processing [7]. While conventional beamforming designs [8] consider antenna selection, they often incur a penalty of beamforming gain. Beamforming gain is crucial in scenarios with multiple antennas [9], and precoding methods can improve the beamforming gain by intelligently adjusting the complex weights on signals at each transmit antenna, ensuring efficient energy transmission to the intended receiver. In contrast to fully digital precoding [10] which activates a large number of antennas simultaneously, hybrid precoding [11] allows for the selective activation of antennas in the analog stage. Only a subset of antennas, determined by the analog precoding strategy, is active at any given time. This selective activation conserves power and enhances energy efficiency. Moreover, a precoding method can adapt a transmission to changing channel conditions by dynamically choosing antennas with better link quality based on real-time measurements. This adaptability improves the system performance under time-varying scenarios.

There are a range of works focusing on precoding schemes [12]-[19]. In [12], a randomized two-timescale hybrid precoding (RTHP) method was developed for the downlink multicell multiple-input multiple-output (MIMO) communications. The proposed algorithm uses a time-sharing mechanism among multiple sets of analog precoders to obtain a tradeoff between fairness and throughput. The complexity of solving such optimization problems may limit the practical feasibility of implementing RTHP in real-time systems, especially in scenarios with fast-changing channel conditions. The authors of [13] proposed a hybrid precoding/combining scheme for mmWave systems in which both partially connected sub-array and fully connected array are investigated. Slow convergence may impact the adaptability of the scheme to dynamic network conditions. In [14], a precoder/combiner neural network architecture was developed which optimizes the precoder/combiner for maximizing the spectral efficiency with imperfect channel state information (CSI) and hardware limitation. Since achieving optimal precoding in wireless communications depends on channel knowledge, the development of a rapid and accurate channel estimation method

A. Pourkabirian is with the School of Computer Science and Statistics, Trinity College Dublin, Dublin, Ireland. E-mail: a.pourkabirian@tcd.ie.
K. Li is with the Real-Time and Embedded Computing Systems Research Centre (CISTER), 4249015 Porto, Portugal. E-mail:kai@isep.ipp.pt.
P. A. Stavrou is with the Communication Systems Department, EURECOM, Sophia-Antipolis, 06410, France. E-mail: fotios.stavrou@eurecom.fr.
W. Ni is with Data61, CSIRO, Sydney, NSW 2122, Australia, and the School of Computer Science and Engineering, the University of New South Wales, Kengston, NSW 2073, Australia. E-mail: wei.ni@ieee.org.

[15] becomes crucial. Most existing studies, e.g., [16],[17] have developed algorithms based on the knowledge of full CSI. While full CSI encompasses detailed knowledge of both magnitude and phase information for all antennas in the array, obtaining full CSI is often challenging because of the time-varying nature of the channel, feedback overhead, and estimation error [18].

In contrast, direction-of-arrival (DoA) estimation, e.g., [19]-[21] focuses solely on determining the direction of maximum signal power from the transmitter antenna. Transmitting pilot signals for DoA estimation consumes less bandwidth and time compared to transmitting signals for full CSI estimation. This can result in lower overhead and more efficient use of resources, which is more practical in certain real-world scenarios. The most common practice of DoA estimation [22] is multiple signal classification (MUSIC), which exploits the spatial signatures of signals received by an antenna array to estimate DoA. Nevertheless, The MUSIC algorithm performs well when the number of received paths is known and not too high. Beam sweeping [23] is another high-resolution technique involving transmitting pilot signals from the antenna array in various directions. This process involves analyzing the received signals, amplitudes, and phases [24],[25]. Beam sweeping is simple but may be slower and less accurate in rapidly changing environments. By identifying the time when the received power reaches its peak, it becomes possible to determine the correct pair of beams and estimate the angle of departure (AoD)/ angle of arrival (AoA) [26] for each propagation path. Estimation of signal parameters via rotational invariance techniques (ESPRIT) [27] is another widely recognized technique, which employs the eigenvalues of the covariance matrix of received signals to accurately estimate DoA, especially when the number of signals is small. Maximum likelihood estimation (MLE) [28] and least squares (LS) estimation [29] are commonly used techniques for DoA estimation. Recently, deep learning methods [30] have been applied to DoA estimation. Neural networks [31] can learn complex relationships between received signals and DoA. Authors of [32] developed a joint channel estimation and hybrid precoding for multi-user millimeter wave massive multi-input multi-output system. However, learning-based methods typically require large datasets for effective training, which might not always be readily available in some practical scenarios such as real-time processing and low-latency responses in virtual reality systems. The study in [33] designed an energy efficiency maximization scheme by jointly optimizing the active phase shifters, RF chain-to-antenna connections, and precoding matrix.

In practical wireless communication scenarios, a complex interrelation [34] exists between the angles at which beams are steered and the phases of signals transmitted by each antenna. As the angle changes, the phase also changes, and these variations are inherently correlated. The correlation between angle and phase is essential for accurately determining the DoA of signals. The phase information is used to estimate the angles of incoming signals, and these estimates are inherently correlated. Conventional approaches commonly treat angles independently, neglecting their potential correlations, which can result in suboptimal performance.

In this paper, we examine the joint statistical behavior of angle and phase components by modeling them with a bivariate Gaussian distribution [35]. Bivariate distribution characterizes the joint behavior of angle and phase random variables, capturing their simultaneous variations and interdependencies. Compare to prior works in the literature, we define a new joint angle and phase entropy to quantify uncertainty or randomness. High entropy [36] indicates that the signals departing from a wide range of directions imply stronger diversity in the propagation paths. The signals can achieve spatial diversity and mitigate fading effects, however, experience independent fading. In contrast, low entropy shows that the signals depart from a limited set of directions, resulting in fewer diverse propagation paths, which is desirable in precoding scenarios. Neglecting the entropy can yield solutions that do not fully utilize the potential benefits of spatial diversity and are sensitive to fading in specific directions.

The key contributions of this paper are summarized as follows.
- We design a hybrid precoder consisting of both analog and digital precoding components for MU-MMIMO systems over fading channels. Using eigenvalue decomposition (EVD), the optimal RF analog precoding matrix is obtained, which maximizes the average sum-rate across all users. We then compute the optimal digital baseband precoding matrix based on minimum mean square error (MMSE) to strike a balance between signal power maximization and interference suppression.
- We consider the inherent interrelation between the angle and phase of a wireless channel and statistically model this correlation as a bivariate Gaussian distribution. This model provides a realistic representation of the angle and phase variations, which is conducive to accurate channel estimation under time-varying channel conditions.
- We define a joint angle and phase entropy to quantify the uncertainty. The proposed precoding approach utilizes joint entropy to dynamically adjust precoding weights in real-time, ensuring optimal signal transmission in dynamic variations.
- We corroborate our theoretical findings with numerical simulations, showing the significant improvement of the proposed approach achieved in estimation accuracy and sum-rate.

The remainder of the paper is organized as follows: Section II describes the system model. In Section III, the joint angle and phase entropy is defined based on bivariate Gaussian distribution. We develop a robust precoding approach for sum-rate maximization and interference mitigation in Section IV. Section V designs receiver strategies for MU-MMIMO systems to establish the complete multi-user equivalent baseband channel between the transmitter and receiver. In Section VI, we develop the robust precoding algorithm to obtain the optimal precoding matrices. The performance evaluation is provided in Section VII. The concluding remarks are provided in Section VII.

*Notation*: Boldface lowercase $\boldsymbol{a}$ and uppercase letters $\boldsymbol{A}$ indicate vectors and matrices, respectively. $\boldsymbol{I}$ represents the $N \times N$ identity matrix, $\mathbb{E}(\cdot)$ signifies statistical expectation, and $\|\cdot\|_F$ denotes the Frobenius norm. The notation used is listed in Table 1.

## II. SYSTEM DESCRIPTION

### A. System Characteristics

Consider an MU-MMIMO system consisting of a transmitter equipped with $N_T$ antennas and $N_{RF} \leq N_T$ RF chains communicates with $K \leq N_{RF}$ users each equipped with $N_R$ antennas. Let $\boldsymbol{F}_{RF} \in \mathbb{C}^{N_T \times N_{RF}}$ be the analog precoding matrix representing the phase adjustments applied to the signal, $\boldsymbol{F}_{BB} \in \mathbb{C}^{K \times N_T}$ be the digital baseband precoding matrix applying complex weights to data streams. The hybrid precoding matrix $\boldsymbol{F} \in \mathbb{C}^{K \times N_{RF}}$ is the product of the analog precoding matrix and the digital precoding matrix $\boldsymbol{F} = \boldsymbol{F}_{RF} \cdot \boldsymbol{F}_{BB}$. This combined matrix represents the complete precoding operation, shaping the signal both in the analog and digital domains.

**Table 1.** Notation and definition

| Notation | Definition |
| --- | --- |
| $N_T$ | The number of transmitter antennas |
| $N_{RF}$ | The number of RF chains |
| $N_P$ | The number of propagation rays |
| $K$ | The number of users |
| $P^{RF}$ | The transmit power of the RF source |
| $\boldsymbol{F}_{RF} \in \mathbb{C}^{N_T \times N_{RF}}$ | The analog precoding matrix |
| $\boldsymbol{F}_{BB} \in \mathbb{C}^{K \times N_T}$ | The digital baseband precoding matrix |
| $\boldsymbol{F} \in \mathbb{C}^{K \times N_{RF}}$ | The hybrid precoding matrix |
| $\mathbf{x} \in \mathbb{C}^{K \times 1}$ | The vector of transmitted signals |
| $\mathbf{s} \in \mathbb{C}^{N_{RF} \times 1}$ | The vector of data streams |
| $\mathbf{y}_k \in \mathbb{C}^{N_R \times 1}$ | The received signal vector of the $k$-th user |
| $\boldsymbol{H}_k \in \mathbb{C}^{N_R \times N_{RF}}$ | The channel matrix for user $k$ |
| $\boldsymbol{P}^{tr} \in \mathbb{C}^{N_{RF} \times K}$ | The matrix of the total transmit power |
| $\boldsymbol{n}_k$ | The vector of additive Gaussian noise |
| $r_k$ | The received data rate at a typical user $k$ |
| $\alpha_n$ | The complex gain of the $n$-th propagation path |
| $\boldsymbol{a}_k(\theta_{n,k}, \varphi_{n,k})$ | The array response vector |
| $\theta$ | The azimuth angle |
| $\varphi$ | The phase |
| $\mu_\theta$ and $\mu_\varphi$ | The means of the angle and phase |
| $\sigma_\theta^2$ and $\sigma_\varphi^2$ | The variances of the angle and phase |
| $\rho$ | The correlation coefficient between angle and phase |
| $\hat{R}(\theta, \varphi)$ | The sample covariance |
| $S(\theta, \varphi)$ | The joint entropy of $\theta$ and $\varphi$ |
| $\boldsymbol{W}_{RF}$ | The RF combining matrix |

Consider the vector of transmitted signals, denoted as $\mathbf{x} \in \mathbb{C}^{K \times 1}$, which is the product of the precoding matrix $\boldsymbol{F}$ and the vector of data streams $\mathbf{s} \in \mathbb{C}^{N_{RF} \times 1}$ ( $\mathbf{x} = \boldsymbol{F}\mathbf{s}$ ) where $\mathbb{E}[\mathbf{s}\mathbf{s}^H] = \boldsymbol{I}$. The resulting received signal at the $k$-th user can then be expressed as follows:

$$\mathbf{y}_k = \boldsymbol{H}_k \boldsymbol{F}_{RF} \cdot \boldsymbol{F}_{BB} \boldsymbol{P}^{tr} \mathbf{s} + \boldsymbol{n}_k, \qquad (1)$$

where $\mathbf{y}_k \in \mathbb{C}^{N_R \times 1}$ denotes the received signal vector for the $k$-th user, $\boldsymbol{H}_k \in \mathbb{C}^{N_R \times N_{RF}}$ denotes the channel matrix composed of channel coefficients for the user $k$, $\boldsymbol{P}^{tr} \in \mathbb{C}^{N_{RF} \times K}$ denotes a diagonal matrix of the total transmit power in which each diagonal element signifies the power allocation dedicated to a particular RF chain, and $\boldsymbol{n}_k \sim \mathcal{CN}(\boldsymbol{0}, \sigma^2 \boldsymbol{I})$ is the $N_R \times 1$ vector of additive Gaussian noise that follows the Gaussian distribution with a mean of zero and a variance of $\sigma^2$. $\boldsymbol{I}$ is the identity matrix which, in the context of a Gaussian distribution, indicates that the random variables in the noise vector are uncorrelated.

To prevent signal distortion due to excessive amplitudes, we describe a normalization constraint $|\boldsymbol{F}_{RF}(i,j)| = 1/\sqrt{N_T}$, where $|\boldsymbol{F}_{RF}(i,j)|$ states the magnitude of the element at the $(i,j)$-th element of $\boldsymbol{F}_{RF}$. To ensure fairness in power allocation among users, we use the Frobenius norm for the transmit power matrix as $\|\boldsymbol{P}^{tr}\|_F \leq P^{max}$. This constraint ensures that the magnitude of the power allocation matrix $\boldsymbol{P}^{tr}$ does not exceed a certain power level, which can be interpreted as a fairness constraint. We also normalize $\boldsymbol{F}_{BB}$ by $\|\boldsymbol{F}_{RF}\boldsymbol{F}_{BB}\|_F^2 = KM$ to ensure compliance with a total transmit power constraint, where $M$ is the number of data streams supported by RF chains, and $\|\cdot\|_F$ denotes the Frobenius norm.

In multi-user scenarios, simultaneous electromagnetic waves originating from other transmitters may interfere with the transmitted signal. Our goal is to develop a precoding algorithm allowing to receive the desired signal while effectively mitigating the impact of interference, subsequently maximizing the sum-rate. According to Shannon's formula, the achievable data rate at a typical user $k$ is given as follows:

$$r_k \approx \log\left(1 + \frac{p_k^{tr}|\hat{h}_k f_{RF,k}^* f_{BB,k}^*|^2}{\sum_{m \neq k} p_m^{tr}|\hat{h}_m f_{RF,m}^* f_{BB,m}^*|^2 + \sigma_n^2}\right), \qquad (2)$$

where $|\hat{h}_k|$ is the magnitude of the estimated channel coefficient matrix for user $k$, $f_{RF,k}^*$ and $f_{BB,k}^*$ are the optimal RF and baseband precoding matrices of user $k$, respectively, $\sum_{m \neq k} p_m^{tr}|\hat{h}_m f_{RF,m} f_{BB,m}|^2$ is the aggregated interference from all interfering sources, and $\sigma_n^2$ is the variance of the additive Gaussian noise.

We aim to maximize the sum-rate of the system to achieve a balance among users and improve the overall system performance. Therefore, we define the following optimization problem as:

$$\max_{\hat{\boldsymbol{h}}_k} \sum_{k=1}^{K} r_k$$
$$s.t.\ C_1: \|\boldsymbol{P}^{tr}\|_F \leq P^{max}$$
$$\phantom{s.t.\ }C_2: |\boldsymbol{F}_{RF}(i,j)| = 1/\sqrt{N_T}. \qquad (3)$$

To compute this optimization problem, we need to find the optimal precoding matrices for both the RF and baseband that maximize system performance while considering power constraints and channel conditions. This optimization process is closely tied to the accurate estimation of channel coefficients since the precoding matrices depend on the knowledge of channel conditions for effective signal shaping.

## B. Channel estimation

We consider a large independent and identically distributed (i.i.d.) Rayleigh fading channel, which is a common consideration in massive MU-MMIMO systems, where all elements of the normalized channel matrix $\mathbf{H}$ are i.i.d. and yield $\mathcal{CN}(\mathbf{0}, \sigma^2 \mathbf{I})$.

***Assumption 1.*** *We consider the uniform propagation environment in which the channel contains a large number of propagation rays $N_P$, densely arranged according to the angle $\theta_n = n\Delta\theta$ in which $\Delta\theta = \frac{2\pi}{N_P}$. Consider both fast and slow time-varying rays of the received signal. Therefore, we define a line-of-sight (LOS) ray, $n_r$ reflected, $n_d$ diffracted, and $n_s$ scattered rays, so that $N_P = 1 + n_r + n_d + n_s$ is the number of propagation paths.*

The channel vector for the $k$ th user is represented as the sum of all scattered propagation paths and is given by
$$\mathbf{h}_k(\theta, \varphi) = \sum_{n=0}^{N_P} \alpha_{n,k}\, \mathbf{a}_k(\theta_{n,k}, \varphi_{n,k}), \quad (4)$$
where $\mathbf{h}_k$ represents the channel vector as a function of angle $\theta$ and phase $\varphi$; $\alpha_n$ identifies the complex gain of the $n$-th propagation path; $-\pi \leq \theta_{n,k} \leq \pi$ and $0 \leq \varphi_{n,k} \leq 2\pi$ are the azimuth angle and phase, respectively, which define the DoA of the user $k$'s signal over the $n$-th propagation path; $\mathbf{a}_k(\theta_{n,k}, \varphi_{n,k})$ is the array response vector representing the directional characteristics of an antenna array at angles $\theta$ and with phase $\varphi$ and is expressed as follows:
$$\mathbf{a}_k(\theta_k, \varphi_k) = \begin{pmatrix} e^{j2\pi f d \sin\theta_k \cos\varphi_k} \\ e^{j2\pi f 2d \sin\theta_k \sin\varphi_k} \\ \vdots \\ e^{j2\pi f(N_P-1)d \sin\theta_k \sin\varphi_k} \end{pmatrix}, \quad (5)$$
in which $f$ is the carrier frequency, and $d$ denotes the spacing between adjacent antenna elements. For conciseness, we replace $\theta_k$ and $\varphi_k$ with $\theta$ and $\varphi$ in the rest of the paper.

## C. Correlation Analysis

In precoding design, the phase information in the array response vector plays a crucial role in steering the antenna array's beam in the desired direction and spatial signal processing tasks like DoA estimation. On the other hand, there exists a physical relationship between the phase and angle of signals. For example, in antenna arrays used for precoding, the phase of each antenna element is determined by the angle at which the signal arrives. As the angle changes, the phase also changes, and these variations are inherently correlated. To capture the statistical dependencies between the angle $\theta$ and phase $\varphi$ components of incoming signals, we model them as a bivariate Gaussian distribution meaning that the angle and phase are correlated Gaussian random variables. This model provides a more realistic representation of the received signals leading to more accurate predictions of channel characteristics.

The joint distribution of $\theta$ and $\varphi$ is expressed as
$$\begin{pmatrix} \theta \\ \varphi \end{pmatrix} \sim \mathcal{N}\left( \begin{bmatrix} \mu_\theta \\ \mu_\varphi \end{bmatrix}, \begin{bmatrix} \sigma_\theta^2 & \rho\sigma_\theta\sigma_\varphi \\ \rho\sigma_\theta\sigma_\varphi & \sigma_\varphi^2 \end{bmatrix} \right), \quad (6)$$
where $\mu_\theta$ and $\mu_\varphi$ identify the means of the angle and phase, respectively; $\sigma_\theta^2$ and $\sigma_\varphi^2$ denote the variances of the angle and phase, respectively; $\rho$ is the correlation coefficient between the angle and phase, which determines the strength and direction of the correlation between the angle and phase and is calculated as
$$\rho = \frac{Cov(\theta, \varphi)}{\sigma_\theta \cdot \sigma_\varphi}. \quad (7)$$
A positive $\rho$ indicates a positive correlation, while a negative $\rho$ indicates a negative correlation. The joint probability density function (PDF) of the bivariate Gaussian distribution for these variables is given by
$$f_{\Theta,\Phi}(\theta, \varphi) = \frac{1}{2\pi\sigma_\theta\sigma_\varphi\sqrt{1-\rho^2}} e^{\left( -\frac{1}{2(1-\rho^2)} \left[ \left(\frac{\theta-\mu_\theta}{\sigma_\theta}\right)^2 - 2\rho\left(\frac{\theta-\mu_\theta}{\sigma_\theta}\right)\left(\frac{\varphi-\mu_\varphi}{\sigma_\varphi}\right) + \left(\frac{\varphi-\mu_\varphi}{\sigma_\varphi}\right)^2 \right] \right)}. \quad (8)$$
To specify the parameters of the joint Gaussian distribution, we employ maximum likelihood estimation (MLE) as follows:
$$\hat{\mu}_\theta = \bar{\theta},\ \hat{\sigma}_\theta^2 = \frac{\sum_{i=1}^{N_P}(\theta_i - \bar{\theta})^2}{n-1},$$
$$\hat{\mu}_\varphi = \bar{\varphi},\ \hat{\sigma}_\varphi^2 = \frac{\sum_{i=1}^{N_P}(\varphi_i - \bar{\varphi})^2}{n-1},$$
and
$$\hat{R}(\theta, \varphi) = \frac{\sum_{i=1}^{N_P}(\theta_i - \bar{\theta})(\varphi_i - \bar{\varphi})}{N_P}, \quad (9)$$
where $\hat{R}(\theta, \varphi)$ represents the sample covariance; $\theta_i$ and $\varphi_i$ denote sample observations of the angle and phase, respectively; $\bar{\theta}$ and $\bar{\varphi}$ are the sample means of the angle and phase, respectively.

**Lemma 1.** Let $\Theta$ and $\Phi$ be two bivariate normal random variables. There exists an independent standard normal random variable $Z_1$ and $Z_2$ such that
$$\Theta = \sigma_\theta Z_1 + \mu_\theta \text{ and } \Phi = \sigma_\varphi\left(\rho Z_1 + \sqrt{1-\rho^2}\, Z_2\right) + \mu_\varphi. \quad (10)$$
***Proof.*** See Appendix A. ∎

**Theorem 1.** The conditional expectation and variance of $\Phi$ given $\Theta$ are given by
$$\mathbb{E}(\Phi | \Theta = \theta) = \mu_\Phi + \rho\, \sigma_\Phi \left(\frac{\theta - \mu_\theta}{\sigma_\theta}\right) \quad (11)$$
and
$$Var(\Phi | \Theta = \theta) = (1-\rho^2)\sigma_\Phi^2. \quad (12)$$
***Proof.*** See Appendix B. ∎

### III. ANGLE ENTROPY ANALYSIS FOR ROBUST PRECODING

In this section, we define a concept of joint angle and phase entropy that measures angular uncertainty or disorder within a bivariate Gaussian distribution. In practice, the channel conditions may be influenced by factors such as fading, mobility, or environmental changes. Entropy can characterize the time-varying nature of the channels by quantifying the variations in angle and phase. A high entropy implies that the channel is subject to fluctuations in angle and phase, which can be due to mobility or environmental changes. This information is imperative for adaptive precoding algorithms that dynamically adjust precoding weights based on channel conditions in real time.

***Definition 1.*** *The entropy $S(X)$ of a random variable $X$ with probability density function $f_X(x)$ is expressed as*
$$S(X) = -\int f_X(x) \log f_X(x)\, dx, \quad (13)$$

*where the integral is taken over the entire space of x.*

Then, the joint entropy of two correlated random variables $\theta$ and $\varphi$ is defined as follows:

$$S(\theta, \varphi) = -\int_{-\pi}^{\pi} \int_{0}^{2\pi} f_{\Theta,\Phi}(\theta, \varphi) \log f_{\Theta,\Phi}(\theta, \varphi) \, d\theta \, d\varphi. \quad (14)$$

**Remark 1.** We make the following remarks on the joint entropy:

$$S(\varphi|\theta) = -\sum_i \sum_j p(\varphi_{i,j}|\theta) \log\left(p(\varphi_{i,j}|\theta)\right) \quad (15)$$

$$S(\varphi|\theta) = -\int_{-\pi}^{\pi} \int_{0}^{2\pi} f_{\Theta,\Phi}(\theta, \varphi) \log\left(\frac{f_{\Phi}(\varphi|\theta)}{f_{\Theta}(\theta)}\right) d\theta \, d\varphi \quad (16)$$

$$\int_{-\pi}^{\pi} \int_{0}^{2\pi} f_{\Theta,\Phi}(\theta, \varphi) d\theta \, d\varphi = 1 \quad (17)$$

$$\int_{0}^{2\pi} f_{\Theta,\Phi}(\theta, \varphi) \, d\varphi = f_{\Theta}(\theta) \quad (18)$$

$$\int_{-\pi}^{\pi} f_{\Theta,\Phi}(\theta, \varphi) \, d\theta = f_{\Phi}(\varphi) \quad (19)$$

$$\int_{-\pi}^{\pi} \int_{0}^{2\pi} f_{\Theta,\Phi}(\theta, \varphi)(\theta - \mu_\theta)(\varphi - \mu_\varphi) d\theta \, d\varphi = Cov(\theta, \varphi) \quad (20)$$

We can extract the phase information $\varphi_{i,j}$ of each complex channel gain $H_{i,j}$ using (15).

**Theorem 2.** Consider that $\theta$ and $\varphi$ are two random variables representing angles or phases with a correlation coefficient $\rho$. The joint entropy of $\theta$ and $\varphi$, denoted as $S(\theta, \varphi)$, with a correlation coefficient $\rho$, can be expressed as a function of their individual entropies, $S(\theta)$ and $S(\varphi)$, as well as their correlation coefficient $\rho$ as follows:

$$S(\theta, \varphi) = -2\pi \log(1 - \rho^2) + S(\theta) + S(\varphi) \quad (21)$$

where the term $-2\pi \log(1 - \rho^2)$ accounts for the correlation between $\theta$ and $\varphi$.

***Proof.*** See Appendix C. ∎

In the following, we revisit the precoding design to solve the optimization problem considering the channel uncertainty.

## IV. PRECODING DESIGN

In MU-MMIMO systems, hybrid precoding, encompassing both RF and baseband aspects, plays a pivotal role in optimizing signal transmission and reception, particularly in the presence of hardware constraints and channel uncertainties. The data streams are multiplied by a hybrid precoding matrix to generate the transmit signal. The hybrid precoding matrix comprises the an RF precoding matrix and a digital baseband precoding matrix. This combined matrix represents the complete precoding operation, shaping the signal both in the analog and digital domains. This signal is sent through the RF front-end for further processing and transmission over the antennas. To achieve efficient hybrid precoding, we start by designing the analog RF precoding.

In MU-MIMO systems, optimizing how signals are sent and received is essential, especially when dealing with limitations in hardware and uncertainties in the wireless channel. Hybrid precoding is a key technique in this process, and it involves two crucial components: the radio frequency (RF) precoding matrix and the digital baseband precoding matrix. These matrices work together to shape the signals before they are transmitted. The RF precoding manages the analog aspects, while the digital baseband precoding handles the digital aspects of the signal. This combined operation prepares the signal for transmission through the RF front-end and over the antennas. To make this whole process efficient, it's essential to begin by designing the analog RF precoding, which is a critical step in optimizing the performance of MU-MIMO systems.

### A. RF precoding design

The goal of the analog RF precoding is to find optimal values for the diagonal entries of the RF matrix, denoted as $F_{RF}^*$, that maximize problem (3). Maximizing $\sum_{k=1}^{K} r_k$ is equivalent to maximizing $\sum_{k=1}^{K} p_k^{tr} |\hat{h}_k f_{RF,k}^* f_{BB,k}^*|^2$. Thus, we rewrite problem (3) as

$$\max_{\hat{h}_k} \sum_{k=1}^{K} \log\left(1 + \frac{p_k^{tr} |\hat{h}_k f_{RF,k}^* f_{BB,k}^*|^2}{\sum_{m \neq k} p_m^{tr} |\hat{h}_m f_{RF,m}^* f_{BB,m}^*|^2 + \sigma_n^2}\right) \quad (22)$$

s.t. $C_1$: $\|P^{tr}\|_F \leq P^{max}$

$C_2$: $|F_{RF}(i,j)| = 1/\sqrt{N_T}$.

We first construct a complex channel vector based on the mean complex channel vector $\boldsymbol{\mu}$ as follows:

$$\boldsymbol{\mu} = \begin{bmatrix} \cos \hat{\mu}_\theta \cdot e^{j\hat{\mu}_\varphi} \\ \sin \hat{\mu}_\theta \cdot e^{j\hat{\mu}_\varphi} \end{bmatrix}. \quad (23)$$

We then define the estimated channel vector for user $k$ as:

$$\hat{\boldsymbol{h}}_k = \begin{bmatrix} \cos \theta_k \cdot e^{j\varphi_k} \\ \sin \theta_k \cdot e^{j\varphi_k} \end{bmatrix}. \quad (24)$$

To characterize the spatial correlation between the elements of the channel vector, we obtain the covariance matrix $\boldsymbol{R}$ as

$$\boldsymbol{R}(\boldsymbol{h}, \boldsymbol{h}^H) = \mathbb{E}[(\boldsymbol{h} - \boldsymbol{\mu})(\boldsymbol{h}^H - \boldsymbol{\mu}^H)^T] =$$

$$\mathbb{E}\left[\begin{bmatrix} \cos\theta e^{j\varphi} - \cos\hat{\mu}_\theta e^{j\hat{\mu}_\varphi} \\ \sin\theta e^{j\varphi} - \sin\hat{\mu}_\theta e^{j\hat{\mu}_\varphi} \end{bmatrix} [\cos\theta e^{j\varphi} - \cos\hat{\mu}_\theta e^{j\hat{\mu}_\varphi}]\right]. \quad (25)$$

To obtain the optimal RF precoding matrix, we employ EVD on the covariance matrix $\boldsymbol{R}$, as given by

$$\boldsymbol{R} = \boldsymbol{U}\boldsymbol{V}\boldsymbol{U}^H, \quad (26)$$

where $\boldsymbol{U}$ represents a unitary matrix comprising the dominant eigenvectors of $\boldsymbol{R}$, which correspond to the key spatial directions for signal transmission; $\boldsymbol{V}$ is a diagonal matrix containing the dominant eigenvalues of $\boldsymbol{R}$. These eigenvalues offer valuable insights into the distribution of signal power across the identified spatial directions.

Eq. (26) allows the RF precoding matrix to take advantage of the channel's second-order statistics through the eigenvalues and eigenvectors obtained from the covariance matrix. Therefore, the RF precoding matrix $\boldsymbol{F}_{RF}$ is influenced by second-order statistics, relying on the eigenvectors and eigenvalues derived from the covariance matrix. Consequently, the entire RF precoding matrix can be expressed as

$$\boldsymbol{F}_{RF} = \boldsymbol{U}\boldsymbol{V}^{1/2} = [\boldsymbol{u}_1 \quad \cdots \quad \boldsymbol{u}_{N_T}] \begin{bmatrix} \sqrt{v_1} & 0 & \cdots & 0 \\ 0 & \sqrt{v_2} & \cdots & 0 \\ \vdots & \vdots & \ddots & \vdots \\ 0 & 0 & \cdots & \sqrt{v_{N_T}} \end{bmatrix}^{1/2} \quad (27)$$

where $\boldsymbol{u}_i$ and $\boldsymbol{v}_i$ denote the $i$-th dominant eigenvector and eigenvalue of $\boldsymbol{R}$, respectively.

### B. Baseband Precoding Design

Following the RF precoder design, the determination of the baseband precoder can be achieved through the MMSE

technique. MMSE precoding provides an excellent trade-off between interference mitigation and signal strength [37]. Typically, the MMSE baseband precoding matrix $F_{BB}$ is given by

$$F_{BB} = \beta \, H^H (HH^H + R_n)^{-1}, \tag{28}$$

where $R_n = \mathbb{E}[n_q n_q^H]$ represents the quantization noise covariance matrix in which $n_q \sim \mathcal{CN}(0, R_n)$ is the additive Gaussian quantization noise and uncorrelated with $s$, and $\beta$ represents the normalization factor applied to ensure compliance with the power constraint which is given by

$$\beta = \left(\frac{s}{Tr\{H^H B^H F_{RF}^H F_{RF} B H\}}\right)^{1/2}, \tag{29}$$

where $B = (HH^H + R_n)^{-1}$ is defined for the brevity of notation. This constraint (29) can be on the total transmit power, per-antenna power, or per-user power.

**Theorem 3.** The optimal baseband precoding matrix $F_{BB}$, which maximizes sum-rate across all users subject to the power constraint, is given by:

$$F_{BB} = \beta \, H^H (HH^H + R_n)^{-1}, \tag{30}$$

where $\beta$ is obtained from (29) and $B$ is given by

$$B = (HH^H + R_n)^{-1}. \tag{31}$$

*Proof.* See Appendix D. ∎

In conjunction with hybrid precoding at the transmitter, the design of efficient combiners at the receiver is also crucial for maximizing the system's performance.

## V. RECEIVER-SIDE STRATEGIES FOR MU-MMIMO SYSTEMS

### A. Combiner Design

We utilize the RF precoding matrix $F_{RF}$ at the transmitter and the RF combining matrix $W_{RF}$ at each user to exploit the substantial array gain offered by the massive array of antennas in the MU-MMIMO channel, all while significantly reducing the number of RF chains to be much smaller than the number of antenna elements at both the transmitter and users. Once $F_{RF}$ and $W_{RF}$ are determined, we can establish the complete multi-user equivalent baseband channel $H_{Eq}$ as follows:

$$H_{Eq} = W_{RF}^H \widehat{H} = \begin{bmatrix} w_1^H & 0 & \cdots & 0 \\ 0 & w_2^H & \cdots & 0 \\ \vdots & \vdots & \ddots & \vdots \\ 0 & 0 & \cdots & w_K^H \end{bmatrix} \begin{bmatrix} \widehat{h}_1 \\ \widehat{h}_2 \\ \vdots \\ \widehat{h}_K \end{bmatrix}, \tag{32}$$

where $W_{RF} = F_{RF}(F_{RF}^H \widehat{H}\widehat{H}^H F_{RF} + I)^{-1} F_{RF}^H$. Each entry of $H_{Eq}$ stands for the equivalent channel gain, representing the transmission gain from an RF chain at the transmitter to an RF chain at a typical user. Now, we can rewrite the optimization problem in (22) as

$$\max_{w_{RF,k}} \sum_{k=1}^{K} \log\left(1 + \frac{\left|\widehat{h}_k^H f_{RF,k}^* f_{BB,k}^* w_{RF,k} (\widehat{h}\widehat{h}^H) w_{RF,k}^H (f_{RF,k}^*)^H (f_{BB,k}^*)^H \widehat{h}_k\right|}{\sum_{m\neq k}\left|\widehat{h}_m^H f_{RF,m}^* f_{BB,m}^* w_{RF,m} (\widehat{h}\widehat{h}^H) w_{RF,m}^H (f_{RF,m}^*)^H (f_{BB,m}^*)^H \widehat{h}_m\right| + \sigma_n^2}\right)$$

$$s.t. \;\; Tr\{W_{RF}^H \widehat{H}\widehat{H}^H W_{RF}\} \leq P^{max} \tag{33}$$

where the constraint ensures that the interference power caused by the receiver combining matrix $W_{RF}$ does not exceed a predefined threshold $P^{max}$. It is essential to keep the interference levels within acceptable bounds in a communication system.

The optimization problem in (33) exhibits concavity due to the logarithmic term in the objective function, signifying diminishing returns as the argument of the logarithm increases. This concavity is further confirmed by the second partial derivative $\frac{\partial^2 r}{\partial w_{RF}^2} < 0$. To address this concave problem, we employ the gradient method as it is highly effective in concave problems, assuring convergence to the global minimum and ensuring the efficient and reliable resolution of our optimization task.

**Lemma 2.** The optimization problem in (33) is concave.
*Proof.* See in Appendix E. ∎

As a result, solving the problem involves finding the optimal $W_{RF}$ that maximizes the sum-rate for all users while satisfying the given constraints.

### B. Gradient-based approach for sum-rate maximization

Calculating the gradient of $r$ with respect to $W_{RF}$ involves taking the derivative of $r$ with respect to $W_{RF}$. It typically requires using matrix calculus and taking partial derivatives with respect to the elements of $W_{RF}$ and then arrange them into a matrix. Therefore, we have

$$\vec{\nabla} r(w_{RF,k}) = \left\{\frac{\partial r}{\partial w_{RF,k}}, k = 1, \ldots, K\right\}. \tag{34}$$

This involves applying the chain rule and taking derivatives of the numerator and denominator separately, and then simplifying the expression. The derivative of the numerator is given by:

$$\frac{\partial \left(\left|\widehat{h}_k^H f_{RF,k}^* f_{BB,k}^* w_{RF,k} (\widehat{h}\widehat{h}^H) w_{RF,k}^H (f_{RF,k}^*)^H (f_{BB,k}^*)^H \widehat{h}_k\right|\right)}{\partial w_{RF,k}}. \tag{35}$$

To simplify this, we consider $A = \left|\widehat{h}_k^H f_{RF,k}^* f_{BB,k}^* w_{RF,k} (\widehat{h}\widehat{h}^H) w_{RF,k}^H (f_{RF,k}^*)^H (f_{BB,k}^*)^H \widehat{h}_k\right|$. Using the chain rule for the square, we can write:

$$\frac{\partial A^2}{\partial w_{RF,k}} = \frac{2A \partial A}{\partial w_{RF,k}}. \tag{36}$$

We can find $\frac{\partial A}{\partial w_{RF,k}}$ by taking the derivative with respect to $w_{RF,k}$ as follows:

$$\frac{\partial A}{\partial w_{RF,k}} = \widehat{h}_k^H f_{RF,k}^* f_{BB,k}^* \left(\frac{\partial W_{RF}}{\partial w_{RF,k}}\right) w_{RF,k}^H (f_{RF,k}^*)^H (f_{BB,k}^*)^H \widehat{h}_k. \tag{37}$$

The derivative of $W_{RF}$ with respect to $w_{RF,k}$ is straightforward and yields:

$$\frac{\partial W_{RF}}{\partial w_{RF,k}} = \delta(i,j) \left(f_{RF,k}^*\right)^H \left(f_{BB,k}^*\right)^H \widehat{h}_k, \tag{38}$$

where $\delta(i,j)$ denotes the Kronecker delta, which is 1 if $i = j$ or 0 otherwise. Therefore, we can rewrite (37) as follows:

$$\frac{\partial A}{\partial w_{RF,k}} = \widehat{h}_k^H f_{RF,k}^* f_{BB,k}^* \delta(i,j) \left(f_{RF,k}^*\right)^H \left(f_{BB,k}^*\right)^H \widehat{h}_k. \tag{39}$$

Note, the derivative of the numerator in (33) becomes:

$$A \hat{h}_k^H f_{RF,k}^* f_{BB,k}^* \delta(i,j) (f_{RF,k}^*)^H (f_{BB,k}^*)^H \hat{h}_k \quad (40)$$

which simplifies to:

$$\left( \hat{h}_k^H f_{RF,k}^* f_{BB,k}^* w_{RF,k} (\hat{h}\hat{h}^H) w_{RF,k}^H (f_{RF,k}^*)^H (f_{BB,k}^*)^H \hat{h}_k \right) \delta(i,j) \quad (41)$$

Next, we proceed to calculate the derivative of the denominator:

$$\frac{\partial \left( \sum_{m \neq k} \left| \hat{h}_m^H f_{RF,m}^* f_{BB,m}^* w_{RF,m} (\hat{h}\hat{h}^H) w_{RF,m}^H (f_{RF,m}^*)^H (f_{BB,m}^*)^H \hat{h}_m \right| + \sigma_n^2 \right)}{\partial w_{RF,k}} \quad (42)$$

This part involves differentiating the sum with respect to $w_{RF,k}$, which would involve taking partial derivatives of each term in the sum. Putting it all together, we can combine the derivatives from the numerator and denominator:

$$\frac{\partial r}{\partial w_{RF,k}} = \frac{\left( \left( \hat{h}_k^H f_{RF,k}^* f_{BB,k}^* w_{RF,k} (\hat{h}\hat{h}^H) w_{RF,k}^H (f_{RF,k}^*)^H (f_{BB,k}^*)^H \hat{h}_k \right) \delta(i,j) \right)}{\sum_{m \neq k} \left| \hat{h}_m^H f_{RF,m}^* f_{BB,m}^* w_{RF,m} (\hat{h}\hat{h}^H) w_{RF,m}^H (f_{RF,m}^*)^H (f_{BB,m}^*)^H \hat{h}_m \right| + \sigma_n^2} \quad (43)$$

This is the partial derivative of $r$ with respect to the element $w_{RF,k}$. To find the entire gradient matrix $\nabla r(W_{RF})$, we repeat this process for each element of $W_{RF}$, resulting in a matrix of derivatives.

**Theorem 4.** The sum-rate maximization problem of (33) has a solution that is achieved when the combining matrix $W_{RF}$ is chosen as follows:

$$W_{RF} = F_{RF} (F_{RF}^H \hat{H}\hat{H}^H F_{RF} + I)^{-1} F_{RF}^H \quad (44)$$

where the elements of $F_{RF}$ satisfy $|F_{RF}(i,j)| = 1/\sqrt{N_T}$.

***Proof.*** See Appendix F. ∎

## VI. PRECODING ALGORITHM

In this section, we develop a robust precoding algorithm that obtains the optimal precoding matrices in the presence of channel uncertainties. The pseudo-code of the algorithm is presented in Algorithm 1.

**Algorithm 1.** The proposed Precoding Algorithm

1. **Set** $t = 0$
2. **Input**: the received signal $y_t$, the maximum transmit power $P^{max}$
3. **Output**: Optimal Precoding matrices $F_{RF}$ and $F_{BB}$, and combining matrix $W_{RF}$
4. **For** each RF chain $i$ of the transmitter antenna
5.   **For** each user $k$
6.     **Initialize** $F_{RF}, F_{BB}, W_{RF}, |F_{RF}(i,k)| = 1/\sqrt{N_T}$, and $|P^{tr}(i,k)| \leq P^{max}$
7.     **Repeat**
8.       Estimate channel vector $h_k(\theta, \varphi)$ using Eq.(4)
9.       Calculate $F_{RF}$ by Eq. (27)
10.       Calculate $F_{BB}$ by Eq. (28)
11.       Calculate $W_{RF} = F_{RF}(F_{RF}^H \hat{H}\hat{H}^H F_{RF} + I)^{-1} F_{RF}^H$
12.       $t = t + 1$
13.     **Until** $\sum_{k=1}^{K} \gamma_k$ is maximized
14.   **End for**
15. **End for**

The proposed algorithm has an overall computational complexity of $O(N_T^3)$. It includes an initialization step with a complexity of $O(KN_T)$ for each RF chain and user, as well as a main loop for precoding calculations for each RF chain with a complexity of $O(N_T)$. Convergence is achieved when a specific criterion (i.e., maximizing the sum-rate) is met.

## VII. NUMERICAL RESULTS

In this section, we numerically evaluate the performance of the proposed method and compare the results with two state-of-the-art schemes, the deep learning-based hybrid precoding method called DLP [32] and the decomposition-based precoding scheme named DBP [33] to verify our analysis. DLP trained a neural network for offline channel estimation in simulated environments, followed by channel reconstruction based on dominant beam space entries, and then uses a separate neural network for hybrid precoder design with phase quantization, transitioning from approximate to ideal phase quantization in the deployment phase for analog precoding. DBP developed a decomposition-based method to address the challenging large-scale mix-integer energy optimization problem in the hybrid precoding method.

We perform simulation for 1000 channel realizations, using Monte-Carlo simulations to compare the performance of our method with these two methods. Our network scenario consists of one radio transmitter which is equipped with $N_T = 64$ antenna elements, $N_{RF} = 16$ RF chains and $K = 8$ users each equipped with 2 antennas. All antennas are assumed to be single-polarized. Random angle and phase channel coefficients are generated based on a bivariate normal distribution. The bound on the maximum transmission power is $P^{max} = 35 \, dB$ and the noise variance is set to $\sigma_n^2 = 0.01$ for all channels. The key numerical values used in the simulation setup are summarized in Table 2.

**Table 2.** Simulation parameters

| | |
|---|---|
| Antenna number | $N_T = 64$ |
| Antenna spacing | $\lambda/2$ |
| Distance of array element from the origin | $32 \lambda$ |
| Number of RF chains | 16 |
| Transmitter height | 10m |
| Receiver (UE) height | 1m-2m |
| Number of users | $K = 8$ |
| Transmit power of the transmitter | $12 \leq P^{tr} \leq 35 \, dB$ |
| Multipath components number | $N_p = 6$ |
| Path loss exponent [33] | $\alpha = 0.4$ |
| Noise power density [33] | -174 dBm/Hz |
| Interference power | -14 dB |
| Carrier frequency | 2.45 GHz |

Fig. 1 illustrates the impact of antenna spacing variation between two antennas on the performance of the proposed precoding methods. The results indicate that the choice of antenna spacing in relation to the wavelength ($\lambda$) has a significant impact on the radiation pattern and interference characteristics of the antenna array. When the antenna spacing is less than $\lambda/4$, it tends to reduce the null depth and increase the side-lobe level. This is because closely spaced antennas have a broader radiation pattern and lower directivity. Conversely, an increase in antenna spacing beyond $\lambda$ results in a smaller beamwidth and higher directivity, leading to narrower main lobes. However, for antenna spacings greater than $\lambda$, the main lobe can split into multiple lobes. For example, at spacings of 1.5 $\lambda$ and 2 $\lambda$, the main lobe may split, creating three lobes. Interestingly, when the antenna spacing is very small, such as 0.3 $\lambda$, the side-lobe levels are minimized, and the null depth is maximized. This configuration offers the best ability to suppress interference from unwanted directions.

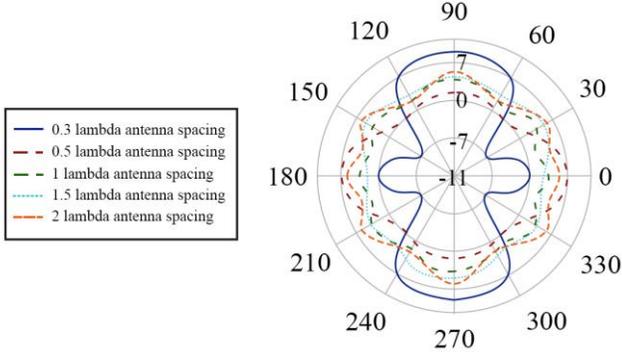

Fig. 1. Impact of antenna spacing on the precoding performance

In Fig. 2, we evaluate the interference power changes with respect to the distance between the transmitter and the users.

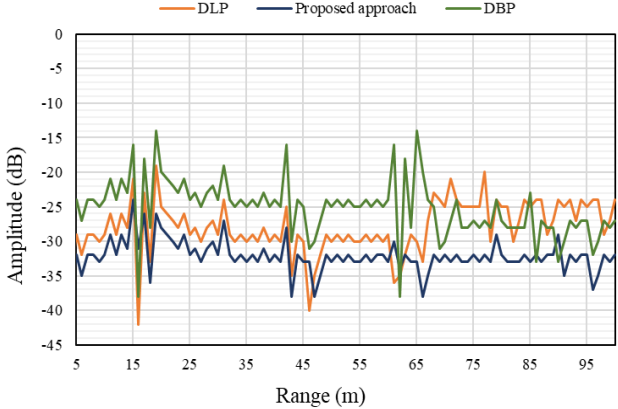

Fig. 2. Received interference power under different distances between the transmitter and the users.

We run experiments for 64 antennas at the transmitter, 16 RF chains, and 6 users which are distributed randomly. We change AoA from $\theta = 12°$ to $\theta = 65°$ and the range between the transmitter and the users. It is shown that the proposed approach has better performance than the two other methods. Our method with the input SNR 30 dB experiences a low side-lobe interference of -33.6 dB, while DLP and DBP schemes are affected by stronger interference of up to -28.2 dB and -24.1 dB, respectively. The reason is that the correlated angle and phase modeling essentially provide a more realistic representation of the physical channel, allowing for better interference cancellation due to exploiting the spatial relationships between signals.

Fig. 3 depicts the sum-rate performance of three schemes where the radio transmitter is equipped with 64 antennas and simultaneously serves $K = 8$ users over the same frequency band in full duplex mode. Three interfering sources with a similar interference-to-noise ratio (INR) of - 14.37 dB affecting the array from different angles $\theta_I = [-7°, 2°, 12°]$ are set. The DoA of the intended signal is 8°. We increase the signal-to-noise ratio (SNR) up to 35 dB and adjust the additive noise power to – 2.8 dB. It is seen that the sum-rate grows as the input SNR of the serving transmitter increases. However, this increase for the two methods DLP and DBP is much less than for the proposed robust algorithm. Compared with the two other approaches, the proposed method achieves the peak sum-rate of up to 1 Gbps, whereas DLP and DBP obtain 0.8 and 0.7 Gbps, respectively. The sum-rate improvement in the proposed approach arises from its ability to accurately model the correlation between angle and phase, leading to better channel characterization and side-lobe interference suppression. Reliance on model assumptions and training data of two other approaches degrades the sum-rate of the systems, especially in time-varying scenarios because the training data does not cover all possible interference scenarios. For instance, the proposed method achieved sum-rate of 2 Gbps under input SNR of 12 dB and INR=-14dB.

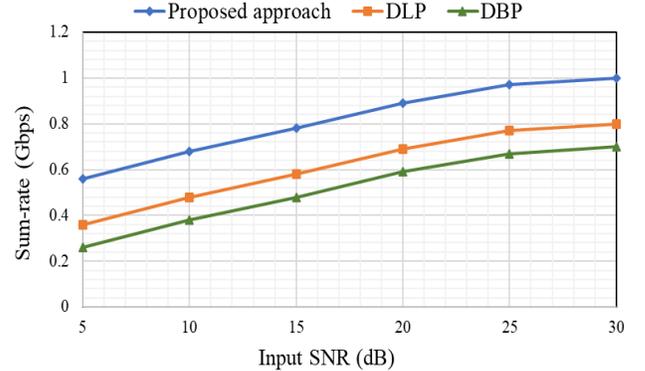

Fig. 3. SINR performance versus input SNR for $K = 8$, INR=-14.37, $\theta_I = [-7°, 2°, 12°]$, and $n = 2.8$ dB.

Fig. 4 shows the worst-case SINR received under different steering angles with $k = 8$ users. We perform the experiments with two different interference powers, i.e., INR= -10 dB and INR= -20 dB, and angular mismatch $\Delta\theta = 7°$. Our results indicate the superiority of the proposed approach, even when the angle mismatch increases gradually because there is always a low interference power due to the robustness against the angular mismatch through the precoding process.

In Fig. 5, we consider another scenario with angular mismatch $\Delta\theta = 13°$ for both INR=-10 dB and INR=-20 dB. The plot shows the SINR performance of all four methods. The results verify the necessity of robust precoding to obtain accurate angle information as increasing the angle mismatch leads to higher side-lobe interference and degrading of the SINR performance.

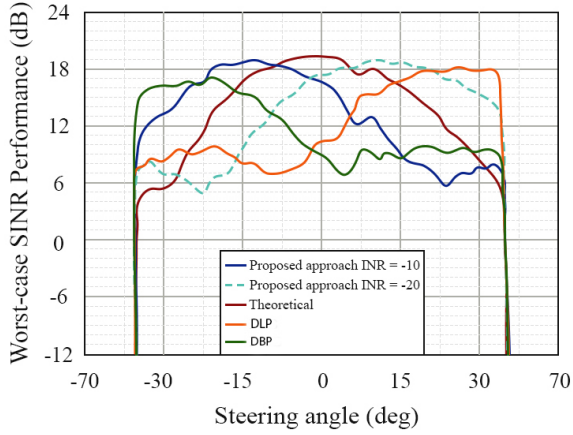

Fig. 4. Worst-case SINR with INR= -10 dB and INR= -20 dB under angular mismatch $\Delta\theta = 7°$.

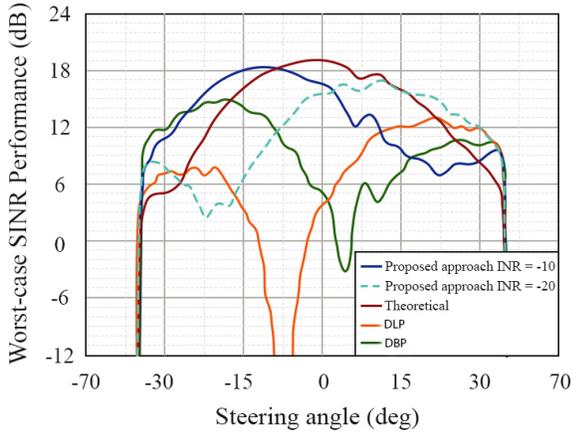

Fig. 5. Worst-case SINR with INR= -10 dB and INR= -20 dB under angular mismatch $\Delta\theta = 13°$.

In Fig. 6, we study the impact of the channel uncertainties which causes the angular mismatch on the bit error rate (BER) versus the number of beams. Three uncertainty scenarios with angle mismatches $\Delta\theta = 5°$, $\Delta\theta = 8°$, and $\Delta\theta = 12°$ are considered. It is observed that the DBP beamformer suffers remarkable BER with even a small angle mismatch $\Delta\theta = 5°$, whereas the proposed approach and DLP are more resilient over this mismatch. For instance, the BER of DBP is $10^{-2}$ when there are eight beams under angle mismatch $\Delta\theta = 8°$. Although the DLP is more robust against the angle mismatch than DBP, its BER is much more than that of the proposed precoder. In other words, the BER of DBP and DLP can only approximate that of the proposed precoder when the angle mismatch is within a limited region. As the angle mismatch rises, the BERs under all methods increase since angle mismatch negatively affects interference suppression. Nevertheless, the BER of the proposed precoder stays relatively the same and does not go over $10^{-6}$.

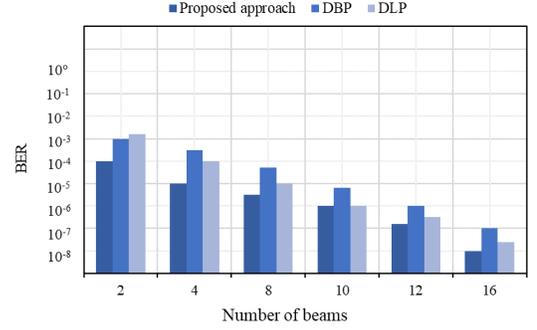

a) angle mismatch $\Delta\theta = 5°$

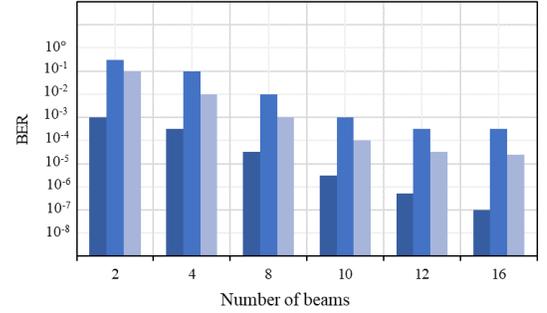

b) angle mismatch $\Delta\theta = 8°$

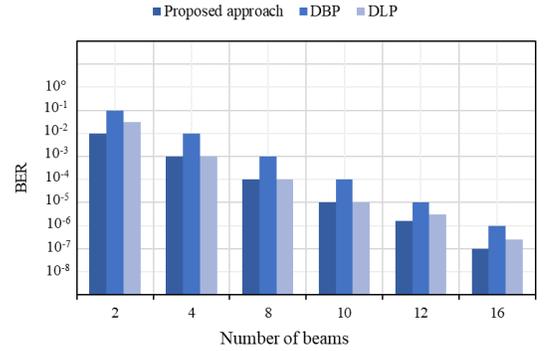

c) angle mismatch $\Delta\theta = 12°$.

Fig. 6. BER versus different numbers of beams under different angle mismatches

Fig. 7 plots the CDFs of the estimation error for all three methods in the case of imperfect channel matrices

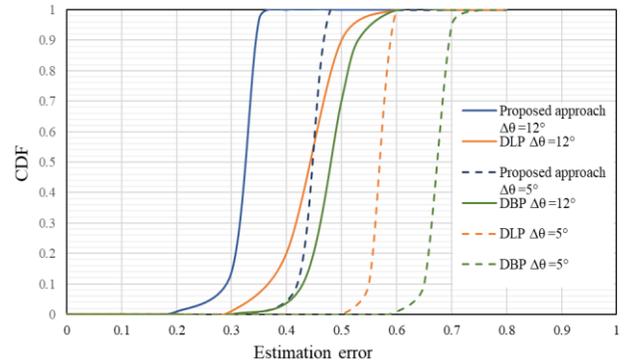

Fig. 7. The CDF of the estimation error of three schemes

## VIII. Conclusion

In this paper, we designed a hybrid precoding approach that leverages both digital and analog precoding techniques to enhance data transmission in MU-MMIMO communications. The proposed approach maximizes the sum-rate of the system while effectively suppressing side-lobe interference by steering signals in specific directions. We identified the inherent correlation between angle and phase in complex signals, modeling them as correlated variables following a bivariate Gaussian distribution. This model allows for a more accurate representation of channel characteristics where precise knowledge of the channel is critical for optimizing signal transmission. The introduction of a joint angle and phase entropy, another key innovation in this research, provided a crucial metric for quantifying the uncertainties associated with angle and phase variations in wireless channels. Through extensive simulations, the paper underscores the robustness and sum-rate maximization achieved by the proposed approach. As a future study, we will employ the phase components to control the time delays and phase shifts applied to signals at each antenna element to achieve desired spatial characteristics.

## Appendix A
## Proof of Lemma 1

Let $Z_1$ and $Z_2$ be two random variables as follows:
$$Z_1 = \frac{\Theta - \sigma_\theta}{\sigma_\theta} \tag{45}$$
and
$$Z_2 = -\frac{\rho}{\sqrt{1-\rho^2}}\left(\frac{\Theta - \sigma_\theta}{\sigma_\theta}\right) + \frac{1}{\sqrt{1-\rho^2}}\left(\frac{\Phi - \mu_\varphi}{\sigma_\varphi}\right). \tag{46}$$
Then, we define the following expression:
$$h_1(Z_1, Z_2) = \sigma_\theta Z_1 + \mu_\theta \tag{47}$$
and
$$h_2(Z_1, Z_2) = \sigma_\varphi(\rho Z_1 + \sqrt{1-\rho^2}\, Z_2) + \mu_\varphi. \tag{48}$$
Using the method of transformations, the joint PDF of $Z_1$ and $Z_2$ can be obtained as follows:
$$f_{Z_1 Z_2}(z_1, z_2) = f_{\Theta\Phi}\big(h_1(Z_1, Z_2), h_2(Z_1, Z_2)\big)\,|J| \tag{49}$$
where $J$ is the Jacobian of h and is expressed as follows:
$$J = \det\begin{bmatrix} \frac{\partial h_1}{\partial z_1} & \frac{\partial h_1}{\partial z_2} \\ \frac{\partial h_2}{\partial z_1} & \frac{\partial h_2}{\partial z_2} \end{bmatrix}. \tag{50}$$
Now, we have:
$$f_{Z_1 Z_2}(z_1, z_2) =$$
$$\frac{1}{2\pi} e^{\left(-\frac{1}{2(1-\rho^2)}\left(z_1^2 - 2\rho(z_1)\left(\rho(z_1)+\sqrt{1-\rho^2}\,z_2\right)+\left(\rho(z_1)+\sqrt{1-\rho^2}\,z_2\right)^2\right)\right)} =$$
$$\frac{1}{2\pi} e^{\left(-\frac{1}{2(1-\rho^2)} z_1^2 + \frac{1}{2(1-\rho^2)}(\rho^2 z_1^2) - \frac{1}{2(1-\rho^2)} z_2^2 + \frac{\rho^2}{2(1-\rho^2)} z_2^2\right)} =$$
$$\frac{1}{\sqrt{2\pi}} e^{-\frac{1}{2} z_1^2} \frac{1}{\sqrt{2\pi}} e^{-\frac{1}{2} z_2^2} = f_{Z_1}(z_1) f_{Z_2}(z_2). \tag{51}$$
As a result, from (45) and (46) we obtain $\sigma_{z_1} = \sigma_{z_2} = 1$, and $\mu_{z_1} = \mu_{z_2} = 0$ that completes the proof. ∎

## Appendix B
## Proof of Theorem 1

From (10), we have:
$$\mathbb{E}(\Phi|\Theta = \theta) = \sigma_\Phi\left(\rho\,\mathbb{E}(Z_1) + \sqrt{1-\rho^2}\,Z_2\right) + \mu_\Phi. \tag{52}$$
Given $\Theta = \theta$, we can obtain the following:
$$Z_1 = \frac{\theta - \mu_\theta}{\sigma_\theta}. \tag{53}$$
As a result, we have:
$$\mathbb{E}(\Phi|\Theta = \theta) = \sigma_\Phi\left(\rho\,\mathbb{E}(Z_1) + 0\right) + \mu_\Phi \tag{54}$$
and
$$Var(\Phi|\Theta = \theta) = (1-\rho^2)\sigma_\Phi^2\, Var(Z_2) = (1-\rho^2)\sigma_\Phi^2. \tag{55}$$
So, the theorem holds. ∎

## Appendix C
## Proof of Theorem 2

First, we start with the definition of joint entropy (14) for the two random variables $\theta$ and $\varphi$ as follows:
$$S(\theta, \varphi) = -\int_{-\pi}^{\pi}\int_0^{2\pi} f_{\Theta,\Phi}(\theta, \varphi).\log f_{\Theta,\Phi}(\theta, \varphi)\,d\theta\,d\varphi \tag{56}$$
where $f_{\Theta,\Phi}(\theta, \varphi)$ is the joint probability density function of $\theta$ and $\varphi$. Given the correlation coefficient $\rho$, the joint probability density function $f_{\Theta,\Phi}(\theta, \varphi)$ can be expressed as follows:
$$f_{\Theta,\Phi}(\theta, \varphi) = f_\Theta(\theta).f_\Phi(\varphi|\theta) = f_\Theta(\theta).f_\Phi(\varphi|\theta, \rho). \tag{57}$$
This decomposition accounts for the correlation $\rho$ between $\theta$ and $\varphi$. Now, the joint entropy $S(\theta, \varphi)$ can be expressed as the sum of three following components:
$$S(\theta, \varphi) =$$
$$-\int_{-\pi}^{\pi}\int_0^{2\pi} f_\Theta(\theta).f_\Phi(\varphi|\theta, \rho).\log[f_\Theta(\theta).f_\Phi(\varphi|\theta, \rho)]\,d\theta\,d\varphi. \tag{58}$$
Using properties of logarithms simplifies the above expression:
$$S(\theta, \varphi) = -\int_{-\pi}^{\pi}\int_0^{2\pi} f_\Theta(\theta).f_\Phi(\varphi|\theta, \rho).[\log f_\Theta(\theta) + \log f_\Phi(\varphi|\theta, \rho)]d\theta\,d\varphi. \tag{59}$$
Expanding the expression and separate the terms involving $\theta$ and $\varphi$ gives:
$$S(\theta, \varphi) = -\int_{-\pi}^{\pi}\int_0^{2\pi} f_\Theta(\theta).f_\Phi(\varphi|\theta, \rho).\log f_\Theta(\theta)\,d\theta\,d\varphi - \int_{-\pi}^{\pi}\int_0^{2\pi} f_\Theta(\theta).f_\Phi(\varphi|\theta, \rho).\log f_\Phi(\varphi|\theta, \rho)\,d\theta\,d\varphi \tag{60}$$
where the first term of the above equation indicates the entropy of $\theta$, and the second term is the conditional entropy of $\varphi$ given $\theta$ with correlation $\rho$. Thus, we can write:
$$S(\theta, \varphi) = -S(\theta) - S(\varphi|\theta, \rho). \tag{61}$$
Using the properties of conditional entropy, we simplify the second term as follows:
$$S(\theta, \varphi) = -S(\theta) - [S(\varphi) - \log(1 - \rho^2)]. \tag{62}$$
Therefore, we obtain (21) and this completes the proof. ∎

## Appendix D
## Proof of Theorem 3

We first rewrite the maximization problem (3) subject to the power constraint as follows:
$$\max_H \gamma = \frac{Tr\{F_{BB}{}^H HH^H F_{BB}\}}{Tr\{F_{BB}{}^H R_n F_{BB}\}}$$
$$s.t.\,Tr\{F_{BB}{}^H B\, F_{BB}\} \leq S \tag{63}$$
where S is defined as the power constraint. Next, we formulate the problem using the Lagrange as follows:

$$L(\pmb{F}_{BB},\lambda) = \frac{Tr\{\pmb{F}_{BB}{}^H \pmb{H}\pmb{H}^H \pmb{F}_{BB}\}}{Tr\{\pmb{F}_{BB}{}^H \pmb{R}_n \pmb{F}_{BB}\}} - \lambda(Tr\{\pmb{F}_{BB}{}^H \pmb{B}\,\pmb{F}_{BB}\} - S) \quad (64)$$

where $\lambda$ is the Lagrange multiplier associated with the power constraint. To find the critical points, we take the derivative of $L(\pmb{F}_{BB},\lambda)$ with respect to $\pmb{F}_{BB}$, and then set it equal to zero, as follows:

$$\frac{\partial L}{\partial \pmb{F}_{BB}} = 0. \quad (65)$$

By solving for $\pmb{F}_{BB}$ at the critical points, and replacing $\left(\frac{S}{Tr\{\pmb{H}^H \pmb{B}^H \pmb{F}_{RF}^H \pmb{F}_{RF} \pmb{B} \pmb{H}\}}\right)^{1/2}$ and $(\pmb{H}\pmb{H}^H + \pmb{R}_n)^{-1}$ with $\beta$ and $\pmb{B}$ of (29) and (31), respectively, we obtain (30). This completes the proof. ∎

## APPENDIX E
### PROOF OF LEMMA 2

We start with the definition of $\alpha_m$ as follows:
$$\alpha_m = \frac{1}{\sum_{m\neq k}\left|\hat{\pmb{h}}_m^H \pmb{f}_{RF,m}^* \pmb{f}_{BB,m}^* \pmb{w}_{RF,m}(\hat{\pmb{h}}\hat{\pmb{h}}^H) \pmb{w}_{RF,m}^H (\pmb{f}_{RF,m}^*)^H (\pmb{f}_{BB,m}^*)^H \hat{\pmb{h}}_m\right| + \sigma_n^2}. \quad (66)$$

Now, we rewrite the optimization problem in (33) as follows:
$$\max_{\pmb{w}_{RF,k}} \sum_{k=1}^{K} \alpha_m \log\left(1 + \left|\hat{\pmb{h}}_k^H \pmb{f}_{RF,k}^* \pmb{f}_{BB,k}^* \pmb{w}_{RF,k}(\hat{\pmb{h}}\hat{\pmb{h}}^H) \pmb{w}_{RF,k}^H (\pmb{f}_{RF,k}^*)^H (\pmb{f}_{BB,k}^*)^H \hat{\pmb{h}}_k\right|\right)$$

s.t. $C_1$:
$$\sum_{m\neq k} \alpha_m \left|\hat{\pmb{h}}_m^H \pmb{f}_{RF,m}^* \pmb{f}_{BB,m}^* \pmb{w}_{RF,m}(\hat{\pmb{h}}\hat{\pmb{h}}^H) \pmb{w}_{RF,m}^H (\pmb{f}_{RF,m}^*)^H (\pmb{f}_{BB,m}^*)^H \hat{\pmb{h}}_m\right| + \alpha_m \sigma_n^2 = 1, m = 1, \dots, K$$
$$C_2: Tr\{\pmb{W}_{RF}^H \hat{\pmb{H}}\hat{\pmb{H}}^H \pmb{W}_{RF}\} \leq P^{max}. \quad (67)$$

Additionally, we define $u_k$ as follows:
$$u_k = 1 + \left|\hat{\pmb{h}}_k^H \pmb{f}_{RF,k}^* \pmb{f}_{BB,k}^* \pmb{w}_{RF,k}(\hat{\pmb{h}}\hat{\pmb{h}}^H) \pmb{w}_{RF,k}^H (\pmb{f}_{RF,k}^*)^H (\pmb{f}_{BB,k}^*)^H \hat{\pmb{h}}_k\right|. \quad (68)$$

This allows us to rewrite the Eq. (67) as follows:
$$\sum_{k=1}^{K} \alpha_m \log u_k = v. \quad (69)$$

Taking the second partial derivative of $v$ with respect to $u_k$ gives us
$$\frac{\partial^2 v}{\partial u_k^2} < 0. \quad (70)$$

This completes the proof. ∎

## APPENDIX F
### PROOF OF THEOREM 4

We start by expressing the Lagrange of the optimization problem (22) as follows:

$$L(\pmb{W}_{RF}, \omega_1, \omega_2) = \sum_{k=1}^{K} \log\left(1 + \frac{\left|\hat{\pmb{h}}_k^H \pmb{f}_{RF,k}^* \pmb{f}_{BB,k}^* \pmb{w}_{RF,k}(\hat{\pmb{h}}\hat{\pmb{h}}^H) \pmb{w}_{RF,k}^H (\pmb{f}_{RF,k}^*)^H (\pmb{f}_{BB,k}^*)^H \hat{\pmb{h}}_k\right|}{\sum_{m\neq k}\left|\hat{\pmb{h}}_m^H \pmb{f}_{RF,m}^* \pmb{f}_{BB,m}^* \pmb{w}_{RF,m}(\hat{\pmb{h}}\hat{\pmb{h}}^H) \pmb{w}_{RF,m}^H (\pmb{f}_{RF,m}^*)^H (\pmb{f}_{BB,m}^*)^H \hat{\pmb{h}}_m\right| + \sigma_n^2}\right) - \omega_1(\|\pmb{P}^{tr}\|_F \leq P^{max}) - \sum_{(i,j)} (i,j)\,\omega_2(i,j)\left(|\pmb{F}_{RF}(i,j)|^2 - 1/N_T\right) \quad (71)$$

where $\omega_1$ and $\omega_2$ are the Lagrange multiplier. Now, we take the derivative of $L(\pmb{W}_{RF}, \omega_1, \omega_2)$ with respect to $\pmb{W}_{RF}$ and set it equal to zero to find the critical points:

$$\frac{\partial L}{\partial \pmb{W}_{RF}} = 0. \quad (72)$$

Solving for $\pmb{W}_{RF}$ at the critical points gives us (44). This completes the proof. ∎